\documentclass[twocolumn]{emulateapj}
\usepackage{apjfonts}
\newcommand{\etal}{{\it et al.} }
\newcommand{\asca}{{\it ASCA} }

\newcommand{\rosat}{{\it ROSAT} }
\newcommand{\iue}{{\it IUE} }

\newcommand{\xmm}{{\it XMM-Newton} }
\newcommand{\chandra}{{\it Chandra} }

\newcommand{\hetg}{{\it HETGS} }

\newcommand{\nela}{Ne~{\sc x}~Ly$\alpha$ }

\newcommand{\oxla}{O~{\sc viii}~Ly$\alpha$ }

\newcommand{\oxlb}{O~{\sc viii}~Ly$\beta$ }

\newcommand{\oxyseven}{O~{\sc vii} }
\newcommand{\oxyeight}{O~{\sc viii} }

\newcommand{\oxysevenr}{O~{\sc vii} (r) }

\newcommand{\neniner}{Ne~{\sc ix} (r) }

\newcommand{\ngc}{NGC~4593 }

\newcommand{\thr}{3C~120 }

\begin{document}

\title{A WARM HOT INTERGALACTIC MEDIUM TOWARDS 3C~120?}

\author{Barry McKernan\altaffilmark{1,2},
Tahir Yaqoob\altaffilmark{2,3}, Richard Mushotzky \altaffilmark{3}, Ian M. George\altaffilmark{3,4}, T. Jane Turner\altaffilmark{3,4}}

%\begin{center}
%{\it Accepted for Publication in the Astrophysical Journal Letters, 16 October 2003}
%\end{center}

\altaffiltext{1}{Present Address: Department of Astronomy, University of Maryland, College Park, MD 20742}
\altaffiltext{2}{Department of Physics and Astronomy,
Johns Hopkins University, Baltimore, MD 21218}
\altaffiltext{3}{Laboratory for High Energy Astrophysics,
NASA/Goddard Space Flight Center, Greenbelt, MD 20771}
\altaffiltext{4}{Joint Center for Astrophysics, University of Maryland,
Baltimore County, 1000 Hilltop Circle, Baltimore, MD 21250}

\begin{abstract}
We observed the Seyfert~I active galaxy/broad line radio galaxy \thr with the \chandra high energy transmission gratings and present an analysis of the soft X-ray spectrum. We identify the strongest absorption feature (detected at $>99.9\%$ confidence) with \oxla (FWHM $=1010^{+295}_{-265}$ km $\rm{s}^{-1}$), blueshifted by $-5500\pm 140$ km $\rm{s}^{-1}$ from systemic velocity. The absorption may be due to missing baryons in warm/hot intergalactic medium (WHIGM) along the line-of-sight to \thr at $z=0.0147 \pm 0.0005$, or it could be intrinsic to the jet of 3C~120. Assuming metallicites of $\sim 0.1 Z_{\odot}$ , we estimate an ionic column density of $N_{\oxyeight}>3.4 \times 10^{16} \rm{cm}^{-2}$ for WHIGM and a filament depth of $<19 h^{-1}_{70}$ Mpc. We find a baryon overdensity $>56$ relative to the critical density of a $\Lambda$-dominated cold dark matter universe, which is in reasonable agreement with WHIGM simulations. We detect, at marginal significance, absorption of \oxla at $z \sim 0$ due to a hot medium in the Local Group. We also detect an unidentified absorption feature at $\sim 0.71$ keV. Absorption features which might be expected along with O~{\sc{viii}} Ly$\alpha$, were not significant statistically. Relative abundances of metals in the WHIGM and local absorbers may therefore be considerably different from solar.

\end{abstract}

%\keywords{galaxies: active --
%galaxies: individual (\thr) -- galaxies: Seyfert --
%techniques: spectroscopic -- X-rays: galaxies -- X-rays: galaxies}

\section{Introduction}
\label{sec:intro}
Around half of the baryons in the universe 
(by mass) should reside in intergalactic filaments of a warm/hot intergalactic medium (WHIGM) 
(Cen \etal 1995, Dav\'{e} \etal 2001). These filaments are believed to be shock-heated to $10^{5} \ \rm{K} < T <10^{7} \ \rm{K}$ 
(see e.g. Cen \& Ostriker, 1999). Most WHIGM may lie at the hotter 
end of this temperature range (Cen \etal 2001), so high spectral resolution 
X-ray detectors such as those aboard \chandra and \xmm are best placed for investigating the `hot' component of the missing baryons. `Hot' WHIGM may recently have been discovered for the first time at X-ray energies (Nicastro \etal 2002; Fang \etal 2002a,b; Rasmussen \etal 2002; McKernan \etal 2003).

We observed the X-ray source \thr with the High Energy Transmission Grating Spectrometer (or HETG--Markert, \etal 1995) and ACIS aboard \chandra. \thr (z=0.033, Michel \& Huchra 1988) is classified both as a Seyfert~1 active galactic nucleus (AGN) and a broad line radio galaxy (BLRG). Here we discuss the serendipitous discovery of an \oxla absorption signature in the \thr spectrum, likely due to WHIGM along the line of sight. 

\section{Observations and Data}
\label{sec:obs}
 
We observed \thr with the \hetg on board \chandra on 2001 December 21
for $\sim 58$~ks, beginning at UT 11:13:52. The \hetg consists 
of two grating assemblies, a High-Energy Grating (HEG) and a 
Medium-Energy Grating (MEG). Only the summed, $\pm1$ order \chandra grating spectra were used in our analysis. We found the counts distribution as a function of cross-dispersion angle could be approximated by a Gaussian model with widths in the range $\sim 0.46-0.49 \arcsec$ for the $\pm 1$ orders of the MEG. This Gaussian model is consistent with a point source, although there is emission of $\sim 6\%$ peak emission at $\sim +3 \arcsec$, possibly due to the extended jet. We accumulated grating spectra along the dispersion direction and within $\sim \pm 3.6 \arcsec$ of the peak in the cross-dispersion direction. The mean MEG and HEG
 total count rates were $0.8134 \pm 0.0038$ and $0.3770 \pm 0.0031$  cts/s respectively. 
We extracted spectra and spectral responses exactly as described in Yaqoob \etal (2003) and obtained 
a net exposure time of 57,224 s (including a deadtime factor of 0.01618). 
During the observation \thr did not vary much; the light-curve yielded an 
excess variance of $0.0043 \pm 0.0007$ (Turner \etal 1999). Since the HEG bandpass only extends down to $\sim 0.8$ keV we use the MEG as the primary instrument. We used $C$-statistic for finding best-fit model parameters, and quote 90$\%$ confidence, one-parameter statistical errors. The harder spectrum, including the Fe-K 
emission line will be discussed elsewhere.

The ACIS CCDs have been undergoing a low-energy QE degradation, due to absorption by contaminants \footnote{http://cxc.harvard.edu/cal/Links/Acis/acis/Cal\_prods/qeDeg/index.html}. There is a model of the contamination pertinent for pure ACIS data (without the gratings). We can use this model as a `worst case'
to estimate the effects on the continuum. Thus, we will give results with and without corrections using
 the ACIS QE degradation model in XSPEC v11.2 ({\tt acisabs}), using default absorption by the contaminants C, H, O, and N in the ratio 10:20:2:1 by number of atoms. 

\begin{figure}[h]
%\vspace{10pt}
%\centerline{\psfig{file=f1.eps,width=8.0in,height=8.0in}
%}
\epsscale{0.8}
\plotone{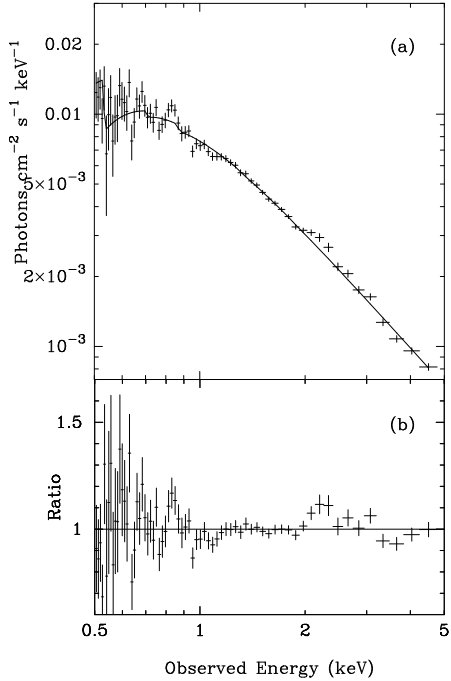}
\caption{Inverted photon spectrum (a), and data/model ratios (b), when 3C~120 
MEG data are fitted with a simple power-law plus Galactic absorption 
model over the 0.5--5~keV band. The 2.--2.5~keV region was omitted during the
fit because large changes in the effective area as function of energy
due to the telescope response, are not pefectly modeled in the response
function. The $\sim 0.8$~keV region also coincides with sharp changes 
in the effective area for this observation, but the magnitude of the
systematic errors is smaller so these data were retained in the fit.}
\end{figure}

All spectral fits were done in the 0.5--5~keV
energy band, excluding the 2.0--2.5 keV region, which
suffers from systematics due to limitations in the calibration of the X-ray telescope
\footnote{http://asc.harvard.edu/udocs/docs/POG/MPOG/node13.html}.
First we fitted spectra binned coarsely at $0.32$A, in order
to compare with previous CCD spectra. A simple model consisting only of a single power--law ($\Gamma = 1.69 \pm 0.02$)
and Galactic absorption of $1.23 \times 10^{21} \ \rm{cm}^{-2}$ (Elvis, Lockman \& Wilkes 1989) fit the data adequately, with residuals $<$ 30 $\%$ over
the fitted energy band.
The inverted photon spectrum and residuals are shown in Fig.~1. The residuals 
at $\sim$ 0.8 keV coincide with sharp changes in effective area and are likely due to the limitations 
of the calibration. Including the ACIS degradation model requires a 
soft excess in the data, which we modeled as a broken power--law. 
We found a best-fit soft X-ray index of $2.58\pm 0.02$, a hard X-ray index 
of $1.76 \pm 0.02$ and a break energy of $1.06^{+0.04}_{-0.06}$ keV. 

We compared our data with previous CCD observations by 
modeling the MEG data with two absorption edges and a single power law
(and Galactic absorption). We find threshold optical depths of $\tau <0.06$ 
and $\tau= 0.06^{+0.05}_{-0.05}$ at $90\%$ confidence for 
\oxyseven and \oxyeight respectively. These small optical depths do not alter the power-law index. These  optical depths agree reasonably well with $\tau <0.01$ (\oxyseven) and $\tau <0.05$ (\oxyeight) found by \asca (Reynolds, 1997). Including ACIS degradation yielded $\tau <0.11$ and $\tau <0.07$ for \oxyseven and \oxyeight respectively. Finer spectral binning ($0.02$A) does not change these results. 

\section{Discrete Spectral Features}
\label{sec:results}

The MEG spectrum of \thr is shown in detail in Fig. 2(a) (0.6--0.8~keV) and Fig. 2(b) (0.8--1.4~keV). Strong absorption appears at $\sim 0.645$ keV and $\sim 0.71$ keV in the observed frame in Fig. 2(a). We identify the feature at $\sim 0.645$ keV with \oxla (0.653 keV rest--frame) blueshifted by $\sim -$ 5500 km $\rm{s}^{-1}$ from \thr systemic velocity. The feature at $\sim 0.71$ keV remains unidentified since all absorption transitions considered for this feature should yield strong absorption features elsewhere in the spectrum. For all spectral fitting hereafter we use a binsize of $0.02$A ( $\approx$ MEG FWHM resolution). 
Since there are bins with zero counts in the \oxla 
feature, the data are of limited statistical quality, modeling the absorption 
profile with a Voigt function is not warranted. Therefore we modeled
the profile with a simple window function, whereby over
an energy interval $W$, centered on some energy, $E$, a fraction
$f$, of the continuum is absorbed.

\begin{figure}[h]
\epsscale{0.8}
\plotone{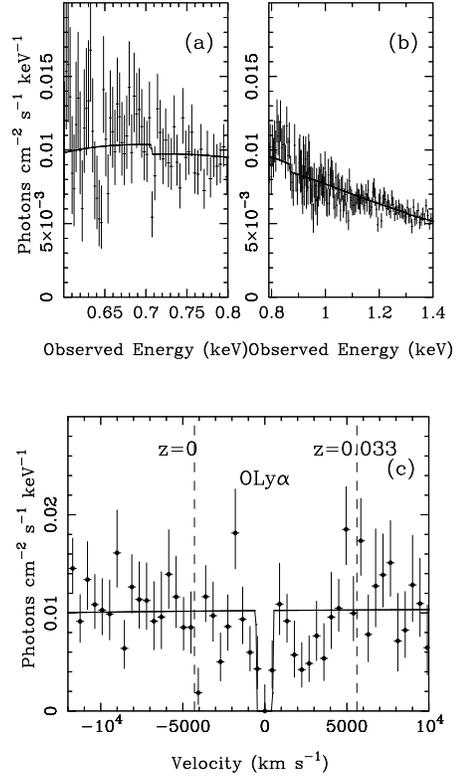}
\caption{(a) \thr MEG observed photon spectrum between 0.6--0.8 keV compared to the best-fitting single power--law model (modified by Galactic absorption). The data are  
binned at $0.08$A. The most significant candidate absorption features in the spectrum lie around $\sim 0.65$ keV and $\sim 0.71$ keV. (b) As for (a) but between 0.8--1.4 keV and the data are binned at $0.04$A between 0.9--1.4 keV. (c) Velocity profile from combined $\pm 1$ order \chandra MEG data centered on the \oxla feature at $0.6443$ keV. A negative velocity indicates a blueshift relative to this energy. Also indicated are the local ($z=0$) and systemic ($z=0.033$) rest-frames. Superimposed is the best--fit absorption line model and continuum (black line). There appear to be two components of \oxla absorption, one marginally significant due to hot medium in the Local Group of galaxies at $z \sim 0$ and the other significant, at $\sim -5500$ km $\rm{s}^{-1}$ blueshifted from \thr (see Table~1). There is no absorption at systemic velocity. Parameters from the model fits are given in Table~1 and can be compared with the FWHM MEG velocity resolution of 360 km $\rm{s}^{-1}$ for \oxla, in the observed energy frame.}
\end{figure}

%\begin{figure}[h]
%\vspace{10pt}
%\centerline{\psfig{file=f2.eps,width=7.0in,height=7.0in,angle=-90}}
%\caption{ }
%\end{figure}

Table~1 shows the results of spectral fitting the \oxla profile. Also in Table~1 are the corresponding results when ACIS degradation is included, and the best--fit to the unidentified feature at $\sim 0.71$ keV (also fit with a window function plus ACIS degradation). The \oxla profile parameters are not strongly influenced by ACIS degradation and the 
C--statistic improves by $>18$ upon the addition of the absorption-line 
model (which corresponds to $>3\sigma$ significance for three additional parameters). The \oxla feature has FWHM$=1010^{+295}_{-265}$ km $\rm{s}^{-1}$, EW$=2.17^{+0.57}_{-0.63}$ eV and is strongly blueshifted ($\Delta v=-5500 \pm 140$ km $\rm{s}^{-1}$) from \thr systemic velocity. The absorption occurs at $z=0.0147 \pm 0.0005$ if it is due to intervening WHIGM. \oxlb absorption is not statistically significant (EW$=0.70^{+1.19}_{-0.70}$ eV) when the data are fit with a Gaussian model with the same FWHM and $z$ as \oxla in Table~1. We tested the data for \oxysevenr, \neniner and \nela absorption in the same way and found EW$<$4.12 eV and $<$1.78 eV for \oxysevenr and \nela respectively. There is weak evidence of \neniner {\it emission} so we could not find a meaningful upper limit on the absorption EW for this feature.

An independent confirmation of the significance of the \oxla feature 
was obtained as follows. From the best-fitting continuum, the instrument
effective area, and the exposure time, we estimated if there 
were {\it no} absorption feature we should expect
a mean of 13.8 photons in the fitted energy interval containing 
the feature (Table~1). We observed 2 photons in this energy interval 
(13 bins out of 126 over the range $z=0$ to $z=0.033$). The Poisson 
probability of obtaining this result is $\sim 0.00975\times (126/13)\%$
so the confidence level of the detection of the feature is
$>99.9\%$ (i.e. $>3\sigma$). We verified this result by performing $10^7$ simulations
of the continuum, with {\it no} absorption feature. 
We note that a 9\% decrease in the level of the continuum is
allowed before the detection significance of the absorption feature
drops below $3\sigma$. Both the \oxla and 0.71 keV absorption features 
are detected separately in the -1 and +1 arms of the MEG. Absorption 
models improved the C--statistic by $17.3,16.0$ for \oxla and by 
$24.8,17.4$ for the $\sim$ 0.71keV feature in the $-1,+1$ arms 
respectively. Thus it is extremely unlikely 
that either feature is due to a statistical fluctuation since the 
probabilities above do not take into account detections in 
\emph{both} arms of the MEG.

Fig.~2~(c) shows a velocity spectrum centered on the \oxla feature. 
Negative velocity indicates blueshift relative to the observed 
energy of the \oxla feature. Superimposed is the best-fitting continuum 
model and absorption line profile as described below (see also Table~1). The 
dashed lines correspond to rest--frame velocities at $z=0$ and $z=0.033$ 
respectively. The \oxla is blueshifted by 
$\sim -5500$ km $\rm{s}^{-1}$ from \thr systemic velocity ($z=0.033$) and there may be absorption at $\sim -4500$ km $\rm{s}^{-1}$ in Fig.~2~(c), close to $z=0$, presumably due to a hot medium in the Local Group, as observed elsewhere (eg Nicastro \etal 2002, Fang \etal 2002b, Rasmussen \etal 2002, McKernan \etal 2003). However, the relativistic jet in \thr could also account for the large outflow velocity (see \S\ref{sec:conclusions}). There is no absorption near the \thr systemic velocity ($z=0.033$). 

\begin{deluxetable}{lrrr}
\tablecaption{Absorption features in the MEG Spectrum of \thr}
\tablecolumns{5}
\tablewidth{0pt}
\tablehead{ &\colhead{\oxla} &\colhead{\oxla} &\colhead{$\sim 0.71$ keV feature} \nl
 &\colhead{ with `{\tt acisabs}'} &\colhead{no correction} &\colhead{ with `{\tt acisabs}'} \nl}
\startdata

$\Delta C$ $^{a}$   &$-$18.5 & $-$18.4 & $-$17.8\nl

$E$ (eV) & $644.3 \pm 0.3$ & $644.3 \pm 0.3$  & $708.5 \pm 0.4$  \nl

$\Delta$ v (km $\rm{s}^{-1}$) $^{b}$ & $-5500 \pm 140$ & $-5500 \pm 140 $ & \ldots \nl

$z_{eff}$ & $0.0147 \pm 0.0005$ & $0.0147 \pm 0.0005$ &\ldots \nl

Width (eV) & $2.17^{+0.57}_{-0.63}$ & $2.20^{+0.52}_{-0.70}$ & $2.00^{+1.40}_{-0.76}$ \nl

FWHM (km $\rm{s}^{-1}$) &$1010^{+295}_{-265}$ & $1025^{+245}_{-325}$ & $930^{+655}_{-355}$\nl

EW (eV) & $2.17^{+0.57}_{-0.78}$ & $2.20^{+0.52}_{-0.92}$ & $2.00^{+1.40}_{-0.54}$ \nl

f$^{c}$ & $1.00^{+\ldots}_{-0.19}$ & $1.00^{+\ldots}_{-0.19}$  & $1.00^{+\ldots}_{-0.38}$  \nl

$\Gamma_{1}$ & $2.58^{+0.10}_{-0.08}$ & \ldots & $2.56^{+0.09}_{-0.07}$ \nl

$\Gamma_{2}$ & $1.76\pm 0.02$ & $1.69 \pm 0.02$ & $1.76\pm 0.02$ \nl

$E_{b}$ (keV)& $1.06^{+0.04}_{-0.06}$ & \ldots & $1.06^{+0.04}_{-0.05}$ \nl

\enddata
\tablecomments{Absorption-line parameters measured
from the MEG spectrum, with (\oxla and $\sim 0.71$ keV feature) and without (\oxla only) the ACIS degradation model. 
The model continuum with ACIS degradation was a broken power--law and the model without 
degradation was a single power--law. 
A  window function was used to
model the \oxla and $\sim 0.7$ keV absorption features,
whereby a fraction $f$ of the continuum is absorbed over an
energy interval, $W$, centered on an energy, $E$ 
(see text for details). All measured quantities refer to the observed frame, 
already corrected for the instrument response.
Errors are 90\% confidence for one interesting parameter
($\Delta C = 2.706$). Velocities have been rounded
to the nearest 5~$\rm km \ s^{-1}$.
$^{a}$ Improvement in C-statistic 
when the indicated absorption feature is added to the
continuum-only model (which includes Galactic absorption).
All values of $\Delta C$ here correspond to $>3\sigma$ significance
for the addition three free parameters.
$^{b}$ Blueshift of feature center energy relative to \oxla in \thr frame. The rest-frame energy of \oxla is 0.6536 keV. $^{c}$ Covering fraction for
the absorption-line model. For a value of $f=1$ 
the width of the feature is equal, by definition, to the equivalent width.}
\end{deluxetable}

\section{Discussion}
\label{sec:conclusions}
We identify the most prominent absorption feature in the soft X-ray spectrum of \thr with blueshifted \oxla ($\sim -5500$ km $\rm{s}^{-1}$ relative to systemic, see Table~1). There may also be absorption at $z=0$ due to \oxla in the Local Group of galaxies and there is unidentified absorption at $\sim 0.71$ keV. The \oxla outflow velocity is considerably larger than in `warm absorbers' in Seyfert~1 AGN (typically a few hundred km $\rm{s}^{-1}$). \oxla absorption possibly due to intervening, non-local WHIGM has been observed in the spectrum of the BL Lac PKS-2155-304 by Fang \etal (2002a) who report a similar width ($<1450$ km $\rm{s}^{-1}$) to our feature ($1010^{+295}_{-265}$ km $\rm{s}^{-1}$), but weaker (EW$=0.48^{+0.25}_{-0.19}$ eV versus EW$=2.17^{+0.57}_{-0.78}$ eV). Absorption by a hot medium in the Local Group ($z \sim 0$) has also been detected towards PKS-2155-304 (Fang \etal 2002a, Nicastro \etal 2002), 3C~273 (Rasmussen \etal 2002, Fang \etal 2002b) and \ngc (McKernan \etal 2003), but these features are generally weaker than the \oxla feature in \thr.

From Table~1, \oxla FWHM$<1305$ km $\rm{s}^{-1}$, so the path length of a putative WHIGM filament has an upper limit $\sim 19 h^{-1}_{70}$ Mpc, otherwise differential Hubble flow would broaden the line. Also, using EW$>1.39$eV and a curve--of--growth analysis, we obtain a column density of $N_{\oxyeight} > 3.4 \times 10^{16} \rm{cm}^{-2}$ (at 90$\%$ confidence). The EW upper limits on O~{\sc{vii}} (r), \oxlb and \nela discussed in \S\ref{sec:results} above, are consistent with this curve--of--growth analysis. Assuming all O is in \oxyeight, and an O abundance of $\sim 0.1$ solar ( or $\sim 8.5 \times 10^{-5} \rm{A}_{H}$), the corresponding neutral Hydrogen column is $\geq 4.0 \times 10^{20} \rm{cm}^{-2}$. For a filament depth $<19 h^{-1}_{70}$ Mpc, we obtain an electron density $n_{e}>6.7 \times 10^{-6} \rm{cm}^{-3}$. If we assume the mean baryon density in the universe ($\Omega_{b}$) relative to the critical density ($\Omega_{\rm{crit}}=9.2 \times 10^{-30} h^{2}_{70}$ g $\rm{cm}^{-3}$) is $\Omega_{b} h^{2}_{70}=0.0224\pm 0.0009$ (Spergel \etal 2003), then using $m_{\rm{H}}=1.7 \times 10^{-24}$ g, the mean number density of baryons ($\bar{N}_{b}$) in the universe is $\bar{N}_{b}=1.2 \times 10^{-7} \rm{cm}^{-3}$. Equating $n_{e}$ with $\bar{N}_{b}$, we find a baryon overdensity ($O_{b}>$56$\Omega_{b} h^{2}_{70}$) in the WHIGM. 
From WHIGM simulations for a variety of $\Lambda$-dominated cold dark matter universes, $O_{b}$ in WHIGM is expected to peak in the range $\sim 10-30$, for distributions where $70-80\%$ of WHIGM baryons lie in the range $O_{b} \sim 5-200$ (Dav\'{e} \etal 2001). Our results indicate that WHIGM may indeed lie in diffuse, hot, intergalactic filaments, slightly denser than predicted by simulations. The high ionization state of the WHIGM filament is consistent with previous observations of \thr in the UV with \iue, which exhibit no obvious Ly$\alpha$ absorption (Kinney \etal 1991). There is also no evidence for absorption in the optical band (Baldwin \etal 1980) in 3C~120. The absence of strong absorption features from \neniner and \nela may indicate that metal abundances in WHIGM are considerably different from solar abundances.

Is it possible that the absorption feature in \thr is actually due 
to intrinsic absorption in the jet? The \oxla feature could originate in an outflow of $\sim 5500$ km $\rm{s}^{-1}$ relative to systemic velocity and many QSOs can exhibit such high velocity outflows. Optically bright QSOs exhibit C~{\sc{iv}} absorption signatures at 5,000--65,000 km $\rm{s}^{-1}$ (Richards \etal 1991). Very broad ($\sim 30,000$ km $\rm{s}^{-1}$) absorption features have been observed in BL-Lacs with \rosat (Madejski \etal 1991), however these are much broader than the features in 3C~120. \xmm observations have found high-velocity outflows in PG1211+143 ($\sim 0.08c$) and PG0844+349 ($\sim 0.2c$) (Pounds \etal 2003a,2003b). If absorption signatures in \thr arise in the jet, they might extend nearly to systemic velocity due to jet acceleration or deceleration. A hint of absorption around $\sim 2500$ km $\rm{s}^{-1}$ in Fig.~2~(c), suggests we cannot rule this out. Interestingly, all the other X-ray sources where WHIGM detection has been claimed (Fang \etal 2002a,b; Nicastro \etal 2002; Rasmussen \etal 2002) possess a jet closely aligned to the line-of-sight. However X-ray sources with jets exhibiting WHIGM absorption may simply represent a selection effect. These are more luminous, distant X-ray sources, so are more likely to have detectable WHIGM filaments along their line-of-sight. 

We acknowledge support from NSF grant AST0205990 (BM), NASA grants NCC-5447 (TY), NAG5-7385 (TJT) and CXO grant GO2-3133X (TY, BM). We used HEASARC online data archive services, supported
by NASA/GSFC and the NASA/IPAC Extragalactic Database (NED) supported by JPL. Thanks to the referee for valuable suggestions to improve this letter.

\end{document}